%% file: main8.tex
\documentclass[
superscriptaddress,
 amsmath,amssymb,
 aps,
pra,
floatfix,
twocolumn
]{revtex4-2}
\usepackage{graphicx}
\usepackage{dcolumn}
\usepackage{bm}
\usepackage[left]{lineno}
\usepackage{xurl}
\usepackage{float}
\usepackage[dvipsnames]{xcolor}

\newcommand{\dcrr}{\text{DCR}_\text{R}}

\begin{document}

\title{Comparing a radiation damage model for avalanche photodiodes through in-situ observation of CubeSat based devices}

\author{Arpad Lenart}
 \email{arpad.lenart@strath.ac.uk}
\affiliation{SUPA Department of Physics, University of Strathclyde, Glasgow G4 0NG, United Kingdom
}%

\author{Tanvirul Islam }%
 \email{cqtmti@nus.edu.sg}
\affiliation{Centre for Quantum Technologies, National University of Singapore, 3 Science Drive 2, Singapore, 117543, Singapore
}%

\author{Srihari Sivasankaran}%
\affiliation{Centre for Quantum Technologies, National University of Singapore, 3 Science Drive 2, Singapore, 117543, Singapore
}%

\author{Peter Neilson}%
\affiliation{Tech-X Corporation, Colorado 80303, United States
}%

\author{Bernhard Hidding}%
\affiliation{SUPA Department of Physics, University of Strathclyde, Glasgow G4 0NG, United Kingdom
}%

\author{Daniel K. L. Oi}%
\affiliation{SUPA Department of Physics, University of Strathclyde, Glasgow G4 0NG, United Kingdom
}%

\author{Alexander Ling}%
\affiliation{Centre for Quantum Technologies, National University of Singapore, 3 Science Drive 2, Singapore, 117543, Singapore
}%
\affiliation{Department of Physics, National University of Singapore, Blk S12, 2 Science Drive 3, Singapore, 117542, Singapore
}%

\begin{abstract}

\textbf{Abstract:} Space-based quantum technologies are essential building blocks for global quantum networks. However, the optoelectronic components used can be susceptible to radiation damage. 
Predicting long-term instrument performance in the presence of radiation remains a challenging part of space missions. 
We present a model that accounts for differences in radiation shielding and can predict the trends for dark count rates of space-based silicon Geiger-mode avalanche photodiodes (GM-APD). 
We find that the predicted trends are correlated with in-situ observations from GM-APDs on-board the SpooQy-1 CubeSat mission.


\end{abstract}

\maketitle

\section*{Introduction}

The democratization of space access, often dubbed New Space, has opened up the potential of low-cost missions utilizing commercial-off-the-shelf (COTS) components. While a rapidly expanding market exists for small satellite platform systems such as CubeSats, on-board scientific instruments may require non-space grade devices and sub-systems. Whilst testing for vibration, thermal, and vacuum conditions are relatively well established, testing for radiation-induced damage remains less representative and amounts to a large source of uncertainty for satellite missions ~\cite{campola2019radiation, cotsuncertainty1, cotsuncertainty2}. Even if a component's radiation tolerance is known, it must be assessed along with measurements or predictions of its exposure to different radiation types, intensity levels, and energy distribution
~\cite{campola2019radiation}.


Space radiation in general occurs due to highly energetic particles like electrons, protons, ions, and neutrons which can damage spacecraft electronics.~\cite{spaceraddamage1,spaceraddamage2,spaceraddamage3}. 
Radiation damage can be categorized as single event effects or cumulative effects. Cumulative radiation damage is further sub-categorized as Total Ionizing Dose (TID) and Displacement Damage Dose (DDD)~\cite{spenvis_background}. 

 Radiation-induced performance degradation is often not well understood for COTS devices which are not radiation hardened by design. Hence the observed degradation may vary significantly on a part-to-part basis ~\cite{campola2019radiation}. Nevertheless, estimating the radiation exposure of instruments and observing the results of radiation tests remain the primary means by which in-orbit performance can be assured.

While radiation models are conventionally used to make component lifetime predictions, we investigate how to use them to predict the gradual degradation of performance due to radiation damage over time. We built a detailed radiation model for the SpooQy-1 cubesat and correlate the results to in-orbit measurements. This radiation model consists of a time-varying,  mission-based radiation fluence predicted by SPENVIS (Space Environment Information System) ~\cite{spenvis_background} combined with a detailed computer-aided design (CAD) model of the SpooQy-1 cubesat.



\begin{figure}[t]
    \centerline{\includegraphics[width=\columnwidth]{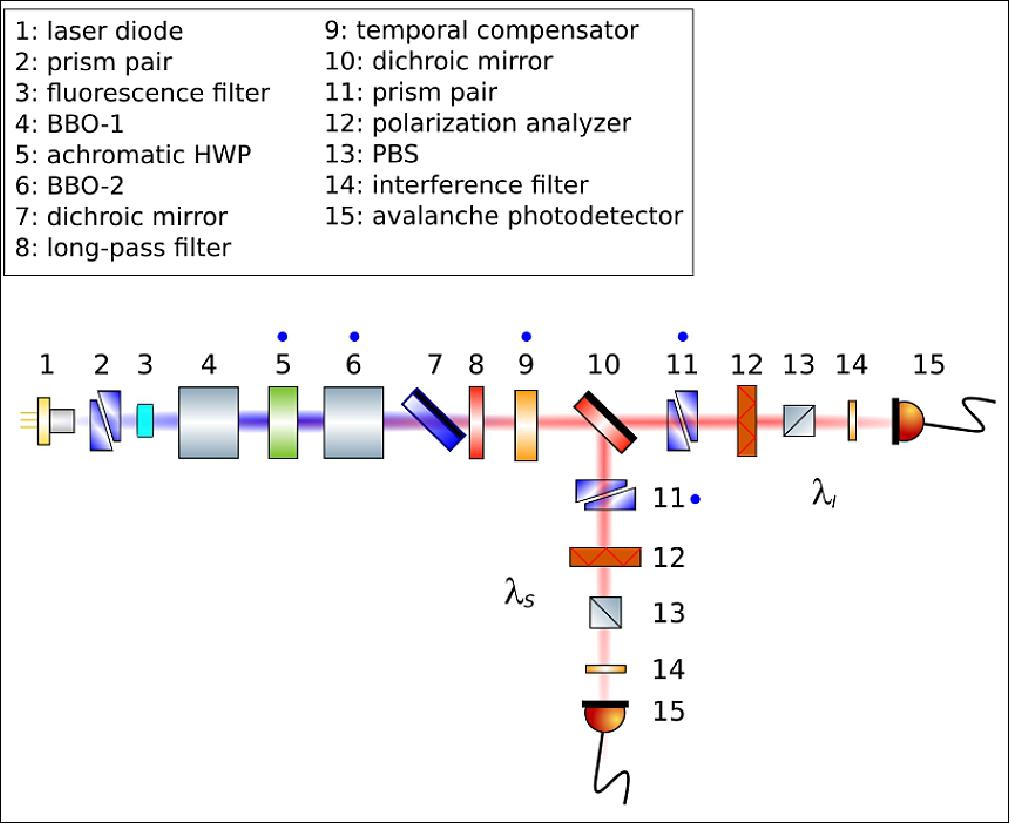}}
    \caption{\textbf{SPEQS-2 Optical layout}: Optical layout of the SPEQS-2 instrument inside the SpooQy-1 satellite. A 405nm laser beam propagates from the laser diode (1) through several optical elements/crystals generating polarization-entangled photon pairs by a collinear, non-degenerate type-I Spontaneous Parametric Down Conversion (SPDC) process~\cite{spooqy1first}. The photons in each pair are separated by a dichroic mirror (10) and detected by separate GM-APDs (15).}
    \label{SPEQS2-SpooQy1-layout}
\end{figure}

SpooQy-1 is a 3U CubeSat deployed into orbit from the International Space Station on 19th June 2019 (de-orbited in late 2021) with operations conducted from Switzerland and Singapore ground stations~\cite{spooqy1first}. The satellite features an entangled photon source and single photon detection system with measurements of polarization-entangled photons performed routinely and successfully since launch. The combined source and detector setup is known as the Small Photon Entangling Quantum System (SPEQS-2 Fig.~\ref{SPEQS2-SpooQy1-layout}), the second iteration of its design. The primary objective of SpooQy-1 was to demonstrate a polarization-entangled photon pair source in space. This would set the pathway for future missions with quantum technologies, towards building global quantum networks. 

SpooQy-1 experienced over 600 days of flight in orbit before de-oribiting. During the operational lifetime of the satellite, the only noticeable change in the components was the increasing rate of background noise in the detectors, attributed to radiation damage. This is of concern because the single photon detectors are free-running, passively-quenched Geiger Mode Avalanche Photodiodes (GM-APDs). All GM-APDs have a recovery time after each photo-detection, during which no other signal can be detected. Increasing background noise will compete with the actual quantum signal leading to a lower signal-to-noise ratio. With sufficient background noise, the devices will become unusable for quantum detection~\cite{yang2019spaceborne}. This is a major motivation for studying how the radiation damage of the GM-APDs can be affected by the internal layout of components.

\begin{figure}[t]
    \centerline{\includegraphics[width=\columnwidth]{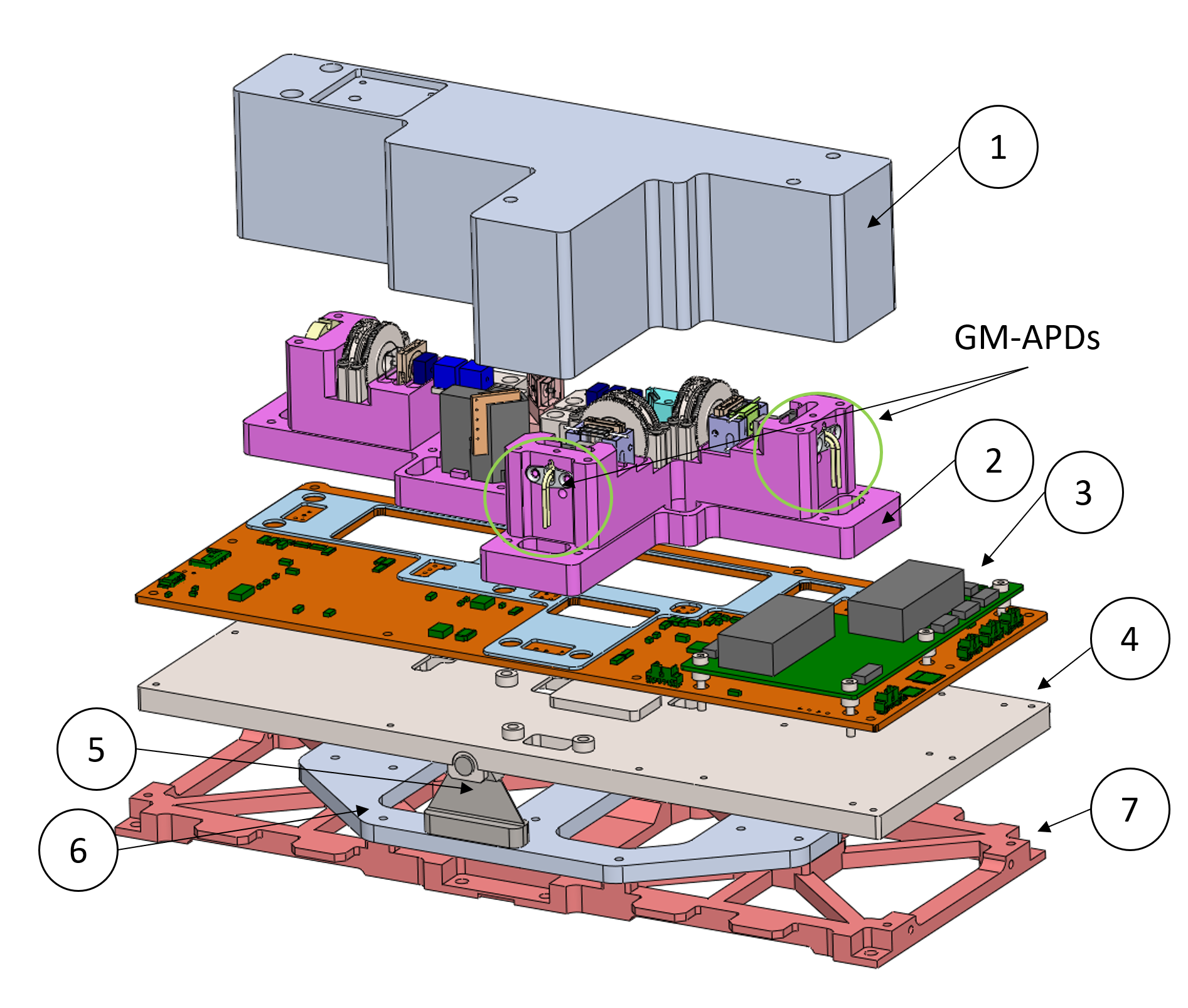}}
    \caption{\textbf{SPEQS-2 exploded view layout}: Mechanical assembly of SPEQS-2 instrument onboard SpooQy-1 satellite: (1) Light tight cover for the optical unit, Material - Aluminium (AL6061-T6). (2) Optical unit (Payload): Single Photon Entangling Quantum System (SPEQS-2), Material - Titanium (Ti-6AI-4V). (3) Onboard electronics mounted onto PCB, Material - FR4). (4) Custom baseplate for the scientific instrument, Material - Titanium (Ti-6AI-4V). (5)Isostatic base mount, Material - Stainless Steel (SS304). (6) Mount base, Material - Aluminum (AL6061-T6). (7) One ribs skeleton, Material - Aluminum (AL7075-T6-2).}
    \label{SPEQS2-SpooQy1-together}
\end{figure}

Radiation modelling requires understanding the mechanical layout acting as radiation shielding materials for critical components. Knowledge of device geometry enables us to assess the effects of differences in shielding materials surrounding the essential components. In some cases, the placement of material can inadvertently cause more radiation damage (for example, by slowing down the protons) Fig.~\ref{SPEQS2-SpooQy1-together} shows the exploded view of the SPEQS-2 assembly~\cite{spooqy1first}, depicting the several layers of mechanical parts of different materials. The optical elements are integrated into the optical unit (Fig.~\ref{SPEQS2-SpooQy1-together}-2), on a custom-made titanium single block. The electronic board is sandwiched between the optical unit and the custom baseplate (Titanium). The optical unit is enclosed with an aluminium cover that ensures the system is light-tight. The isostatic base mount (Stainless steel) and aluminium mount base act as the structural interface between the payload and the skeleton of the satellite.

\begin{figure}[t]
    \centering
    \includegraphics[width=0.95\columnwidth]{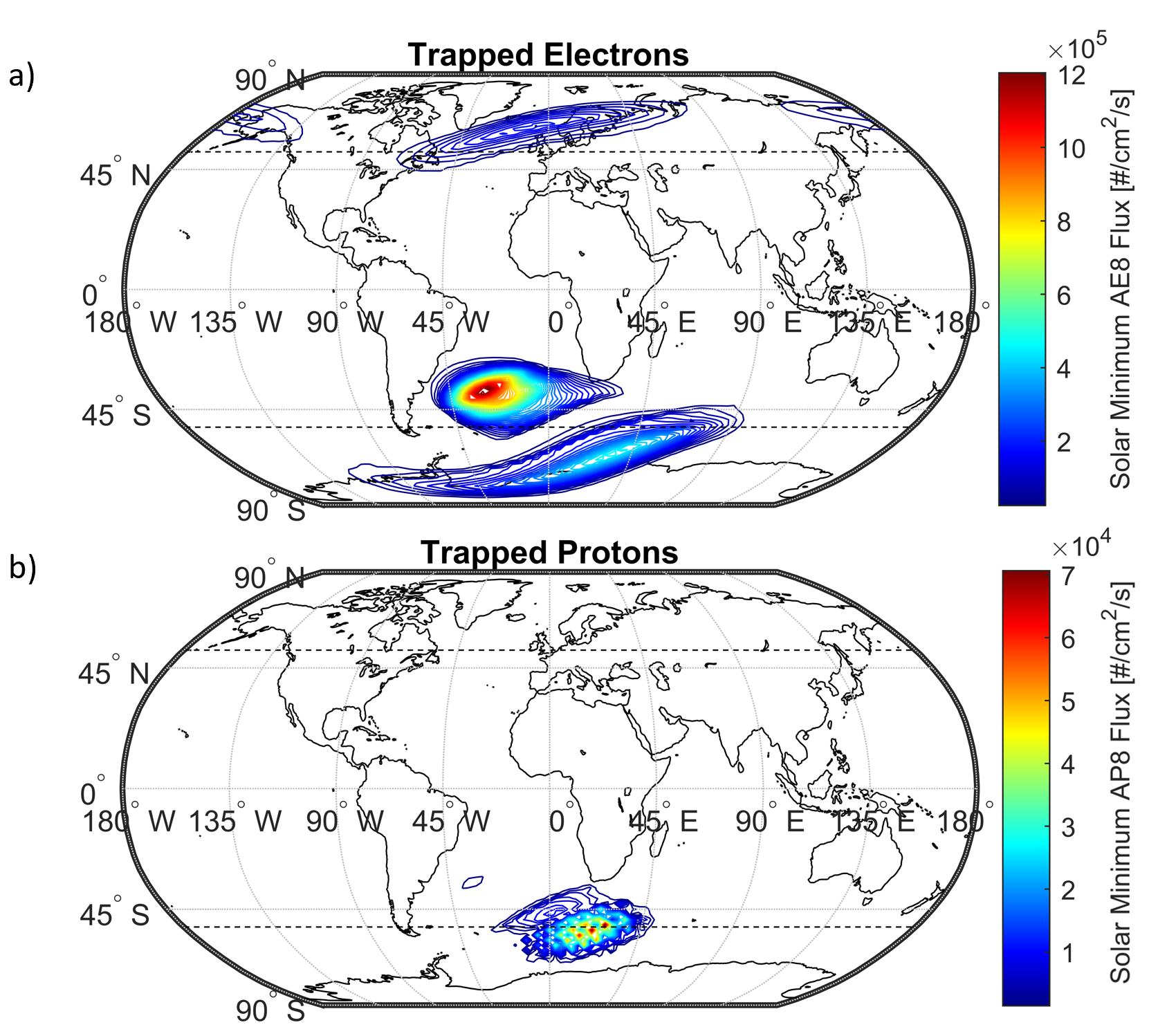}
    \caption{\textbf{Low-Earth Orbit trapped particle fluences:} Abundance of trapped particles regions at 400 km altitude using fluences from SPENVIS. (a) trapped electron population. (b) trapped proton population. The  South Atlantic Anomaly (SAA) South East of South America is where particles are most abundant. The SpooQy-1 orbit at an orbital inclination of $51.6^\circ$ (between the dashed lines) overlaps with the SAA.}
    \label{regions_radiation}
\end{figure}

Satellites face environmental hazards such as launch conditions, vacuum, frequent large temperature cycles, and space radiation. Radiation from energetic particles (electrons, protons, ions, and neutrons) can damage spacecraft electronics and components. Radiation in Low Earth Orbit (Fig.~\ref{regions_radiation}) is mostly due to electrons and protons trapped by the Earth's magnetic field in the Van Allen radiation belt~\cite{spenvis_background}. Their density strongly depends on the inclination and altitude of the satellite's orbit. The orientation of a spacecraft is also important to consider as the flux of space radiation may assume an angular distribution, whereby more radiation can be received from a certain direction due to the East-West effect \cite{spaceraddirectionality,eastwesteffect}. Since CubeSats are constrained by size, weight, and power, it is crucial to understand various materials' radiation shielding capabilities and how space radiation can damage internal instruments.

Charged particles in space radiation deposit energy in materials via several mechanisms, which can be ionizing or non-ionizing for the material and result in their degradation
~\cite{campola2019radiation, bethebloch, NASAprotonguidelines}.
Radiation dose is defined as the amount of energy deposited per unit mass (commonly using rads [0.01 J/kg] or grays [1 J/kg]). Displacement damage is a form of non-ionizing dose resulting in the displacement of atoms from lattice positions. Radiation testing of GM-APDs and previous works ~\cite{yang2019spaceborne, apddcrexponential, cqtspaceborne} reveal they have a very high sensitivity to displacement damage but relatively low sensitivity to ionizing radiation. 

The radiation environments around the Silicon GM-APDs onboard the CubeSat are simulated using Monte Carlo Geant4-based radiation software RSim ~\cite{rsim, geant4}. This model predicts that one of the detectors should experience 1.91 ($\pm$ 0.27) times the amount of radiation exposure as the other. This correlates well with in-situ observation of increased noise in the form of the dark count rate (DCR).

Our analysis in this work indicates that the performance degradation was radiation-induced. While the part-to-part variability in GM-APD behaviour may be a contributor ~\cite{apddddvariability}, the agreement between simulated radiation effects and in-situ measurements suggest that the results are likely due to varied shielding levels arising from the internal CubeSat structure. For CubeSats, such radiation dose inhomogeneity in high fidelity simulations together with corroborating in-orbit observations have not been reported before.

\section*{Results and Discussion}\label{resultsdiscussion}

As our main result, we investigate and correlate the $\dcrr$s in the two GM-APDs with the simulated radiation doses over the duration of the space mission. The details of the radiation model and associated assumptions are discussed in the Methods section.

\subsection*{Radiation Simulation Results}

\begin{figure}[t]
    \centering
    \includegraphics[width=\columnwidth]{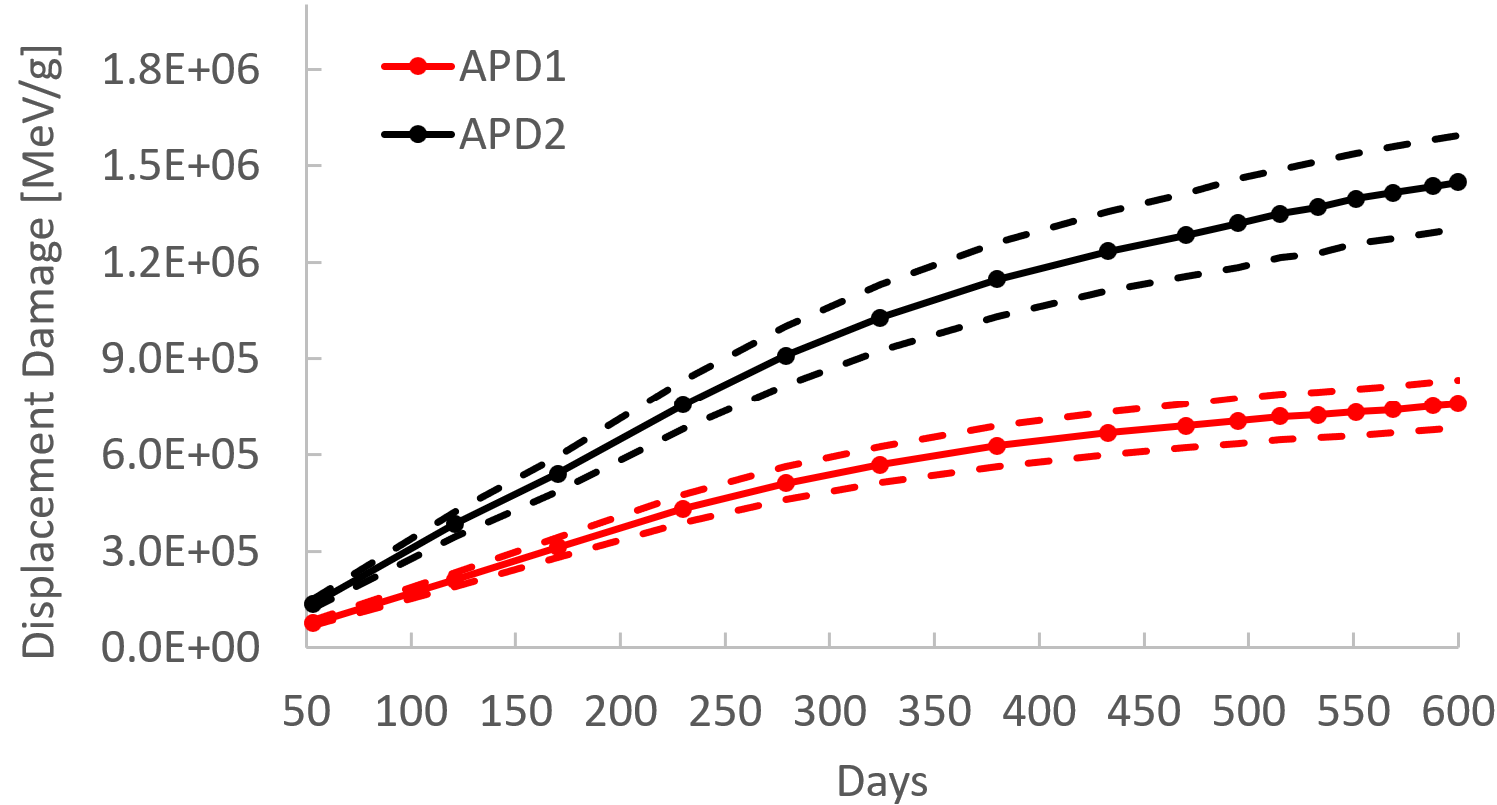}
    \caption{\textbf{Cumulative displacement damage:} The simulated accumulation of displacement damage doses in the GM-APDs over SpooQy-1's lifetime in 2.5 km increments of decreasing altitude. The dashed lines indicate possible values for the radiation doses with maximum and minimum limits (1 standard deviation).}
    \label{TIDDDD}
\end{figure}

\begin{figure}[t]
    \centering
    \includegraphics[width=\columnwidth]{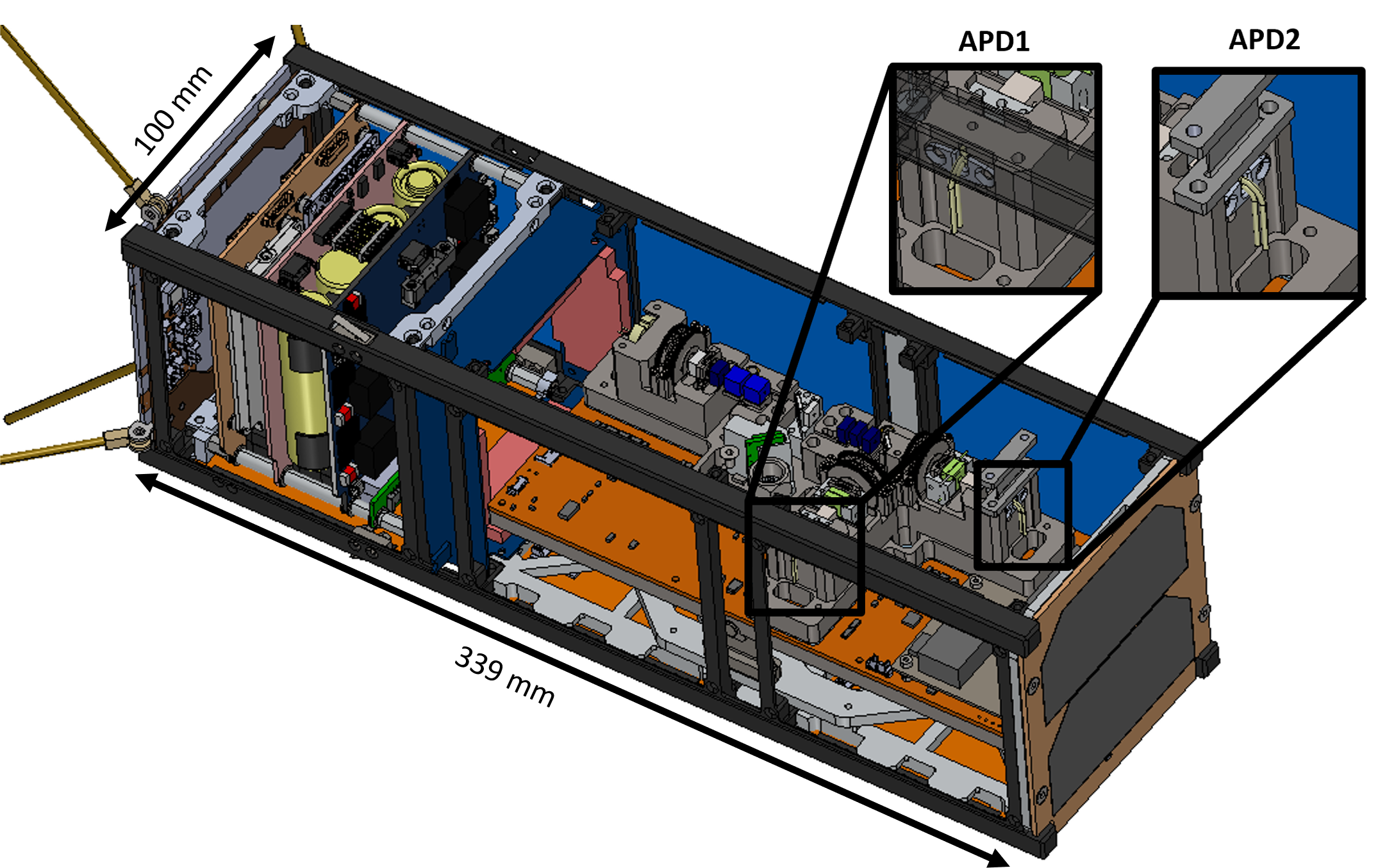}
    \caption{\textbf{SpooQy-1 CAD Model: } Locations of the GM-APDs (APD1 and APD2) onboard SpooQy-1. Various components surround these detectors; some have been hidden here for the reader to understand the GM-APDs' geometric placements. Hidden components include the aluminium cover that ensures the SPEQS payload is light-tight and solar panels.} %
    \label{APD_locations}
\end{figure}

The accumulation of DDD for the two GM-APDs over 600 days is simulated and is shown in Fig. ~\ref{TIDDDD}. The simulations showed that the contributions from solar protons and galactic cosmic ray protons were found to be negligible. The majority of the dose was from trapped electrons and protons. %
The main differences in radiation exposure between the two GM-APDs are observed in primary particles.

\begin{figure*}[ht]
    \centering
    \includegraphics[width=\textwidth]{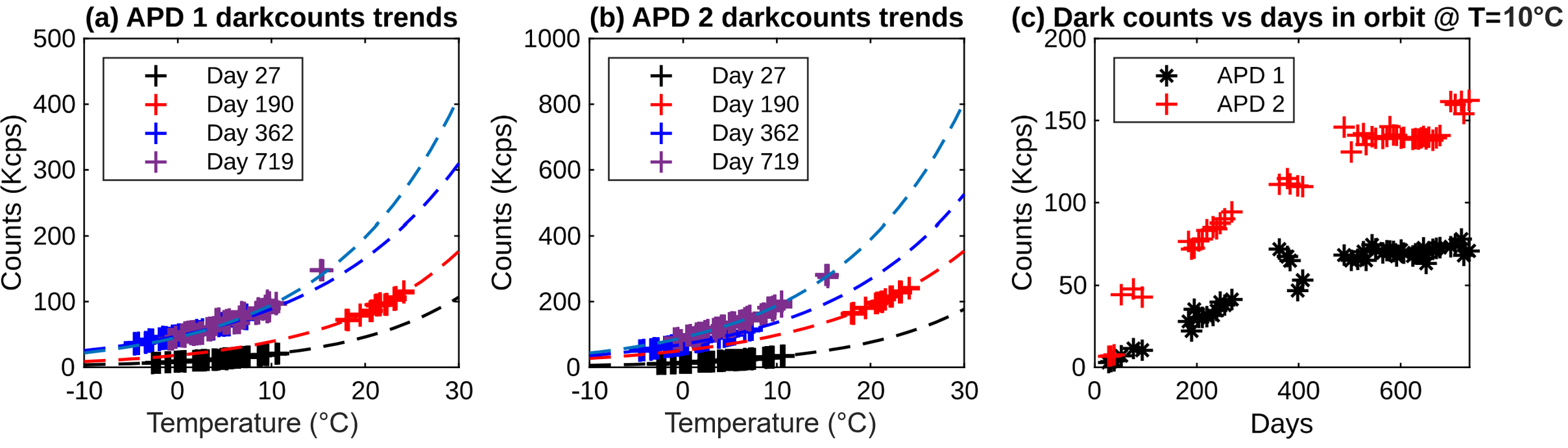}
    \caption{\textbf{Dark count trends: } Dark count rate and temperature measurements directly from the two GM-APDs, APD1 and APD2. On the ground before launch, APD1 and APD2 dark count rates at $20^\circ$C were approximately 17 kcps and 26 kcps respectively. (a),(b) - Dark counts recorded over one orbit plotted as a function of its temperature during the measurement for 27, 190, 362, 719 orbital days. (c) - All of the dark count rate measurements are adjusted to $10^\circ$ C using the exponential fits from (a) and (b). The initial DCR recorded on the ground are subtracted to help capture the radiation induced dark counts ($DCR_R$).}
    \label{DCRtrends}
\end{figure*}

The hypothesis is that there are lower levels of shielding in the regions of GM-APD APD2 (Fig.~\ref{APD_locations}), which results in increased radiation exposure. This is supported by the radiation heatmap (See, Supplementary Figure~S8 and Supplementary Table~I), which is consistent with the trends shown in Fig.~\ref{TIDDDD} of the DDD experienced by each GM-APD. The yearly simulated absorbed radiation dose is expected to be on the order of ~300 rads, as the GM-APDs $\dcrr$ are predominantly affected by displacement damage dose (caused by protons) (with minimal sensitivity to ionizing dose)~\cite{yang2019spaceborne}, the absorbed dose from electrons (mainly ionizing~\cite{spenvis_background}) is expected to have minimal impact on GM-APD performance.

Therefore, based on our simulation model, we expect different levels of radiation damage in the GM-APDs. We reconstructed the evolution of the expected radiation environment by accounting for changes caused by incremental drops in altitude over the duration of the space mission. Based on the assumptions established in the 
Radiation Modelling subsection of the Methods section, one would be able to correlate trends in DDD, to trends in observed dark count rates.

\subsection*{Comparison of Simulation Results to In-Orbit Data}

The dark count rate of GM-APDs increases exponentially with temperature\cite{sap500datasheet, apddcrexponential, apdthermalnoiseexponentialtemperature,anisimova2021low}. 
The observed dark counts for the GM-APDs onboard SpooQy-1 are shown in Fig.~\ref{DCRtrends}(a) and Fig.~\ref{DCRtrends}(b) for various temperatures (on arbitrarily chosen four days as representatives spanning the mission lifetime). Exponential fits are performed on the entire dataset, which are then used to extrapolate dark count rates for a normalized temperature of 10 degrees Celsius, (for illustration, the fit parameters for the representative four days are given in in Supplementary Table~II.) the initial DCR are subtracted for these curves to better compare the increase in $DCR_R$. We note that the operating temperature of the GM-APDs on different days are often quite similar, because operations are typically conducted when the satellite begins to enter the Earth's shadow. This leads to a consistent set of operational data. 

The dark count rates as a function of time are plotted in Fig.~\ref{DCRtrends}(c), and the GM-APD named ``APD2'' has almost twice the dark count rate compared to ``APD1'' by the end of the mission.
The dark count rates increase linearly with orbital days. 
We observe that there are two rates of increase---a relatively steeper rate in the first 500 days, followed by a very shallow rate of increase. This trend is also noticeable in the simulated DDD in Fig.~\ref{TIDDDD}. This drop in dark count rate increment is likely due to less abundant radiation particles and reduction in their energy (specifically protons, the primary contributor to displacement damage) at lower altitude, which causes lower radiation damage.

\begin{figure}[t]
    \centering
    \includegraphics[width=\columnwidth]{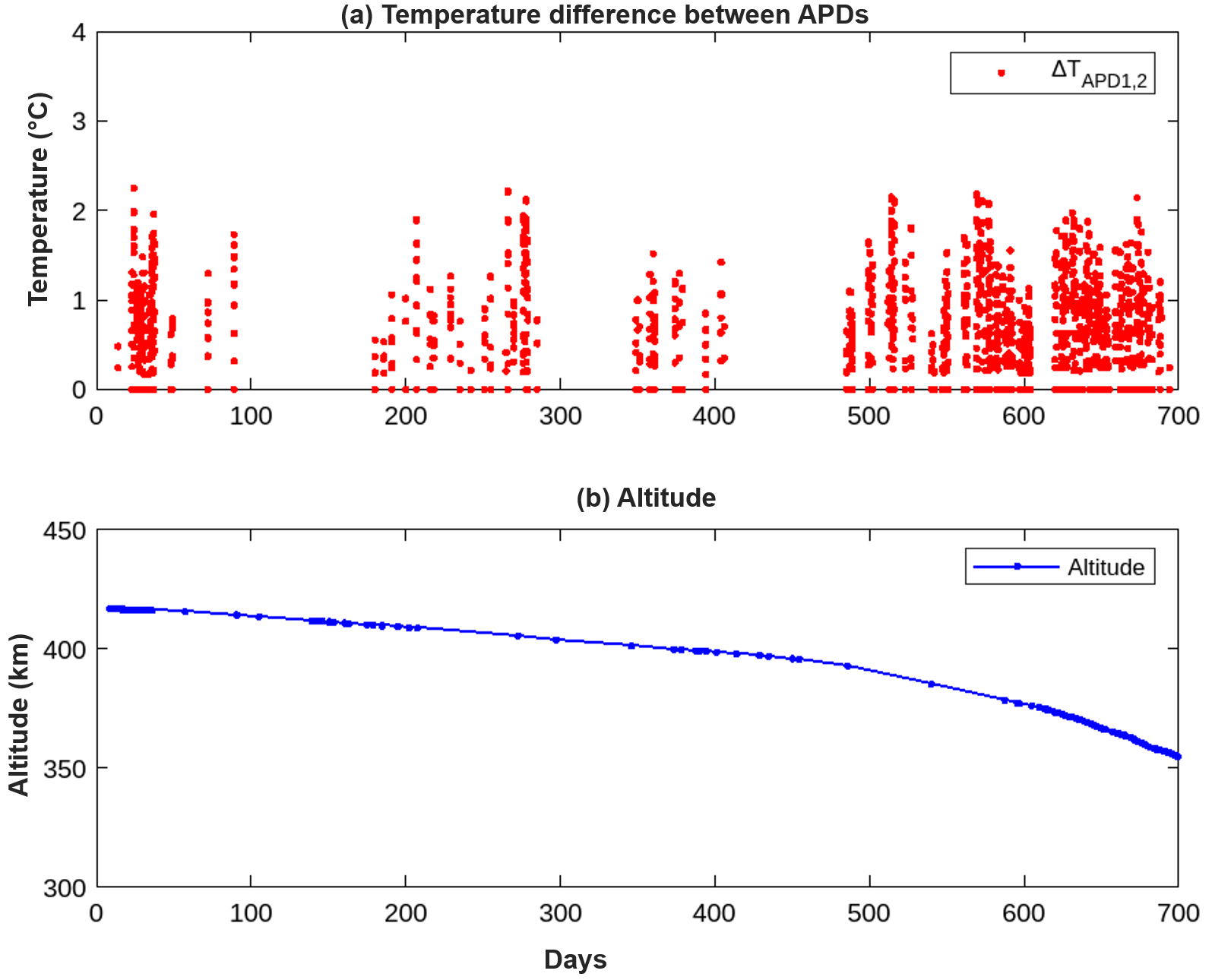}
    \caption{\textbf{SpooQy-1's in-orbit temperature and altitude:} (a) Observed temperature difference between GM-APDs over the mission duration (700 days). This data demonstrates that neither GM-APD experienced large temperature excursions and were relatively close in temperature. There are gaps in the data because the payload was not operated every day. (b) Orbital altitude during mission lifetime.}
    \label{DCR_altitudedata}
\end{figure}

Fig.~\ref{DCR_altitudedata}(a) shows the temperature difference between the two GM-APDs onboard the SPEQS-2 instrument. The average temperature difference is approximately $1^\circ$ C throughout the mission lifetime. This temperature difference is too small to account for the observed difference in the dark count rate between the detectors. For example, consider the data on day 190 of the mission in Fig.~\ref{DCRtrends}. At $20^\circ$C, the difference in dark counts between APD1 and APD2 is approximately 90 kcps. At the same temperature, the dark count rate of APD1 in a temperature range of $6^\circ$C is only between 75 and 110 kcps, or 5.8 kcps per degree C. In the same range, the dark count rate of APD2 is between 180 kcps and 220 kcps, or 6.7 kcps per degree C.

The observed in-orbit dark count rates and simulated displacement damage doses are in general agreement and follow trends from previous reports~\cite{sun01, cqtspaceborne, yang2019spaceborne}.
However, discrepancies due to part-to-part variations~\cite{apddddvariability} cannot be easily modelled, and this prevents a precise prediction of how the dark count rate on each component could increase with radiation damage. Another observation is that the in-orbit dark count rate stopped its rapid increase after about ~500 days, which corresponded to the satellite altitude falling below 400 km, as shown in Fig.~\ref{DCR_altitudedata}(b).
This is also in agreement with lower radiation doses at lower altitudes ~\cite{AltitudeVsDose}.
As the observed count rate is within the linear regime of the in-flight quenching detector circuits, this change in the rate of dark count increase is not due to the detectors becoming saturated with avalanche events.

Whether radiation modelling can be used as a predictor of radiation-induced performance degradation remains an open question. The mechanisms behind part-to-part variability in radiation damage of GM-APDs are not fully understood ~\cite{apddddvariability}. Although we examined the advantages of using realistic radiation models to forecast deterioration patterns in GM-APDs for Low-Earth Orbit Cubesats, it would be worthwhile to investigate whether the radiation environment encountered during spaceflight could be reproduced using ground-based laser-wakefield accelerators ~\cite{laserplasma} that better mimic the energy spectrum than currently employed beamline-tests. Such an approach would narrow the gap between spaceflight and ground-testing conditions and when combined with realistic radiation models may prove to be beneficial in predicting spaceflight performance of future space missions.

\section*{Conclusion}

This study has presented an approach to estimating the radiation dose damage for a critical component through radiation modelling with time-varying fluences and detailed CAD model of a CubeSat.  It is notable that a small volume spacecraft such as a CubeSat can exhibit such a variation in the internal radiation environment and that this was found to correlate with the observed increase in dark counts at the GM-APDs. This detailed modeling technique is expected to become a useful tool for understanding the internal radiation environment inside spacecraft, and be used in the future to optimize in-orbit performance of radiation sensitive components, thus increasing its operational lifetime. This radiation model can be modified for future missions by changing the CAD models and mission fluences.

\section*{Methods}
\label{sec:methods}

\subsection*{Radiation effects on GM-APDs}

To simplify the radiation modelling, we assume that there is no part-to-part variability in the DDD induced degradation across the two GM-APDs. The \emph{radiation-induced} dark count rate $\dcrr$ (counts per second) for a GM-APD can be modelled as~\cite{yang2019spaceborne}, 
\begin{equation}
    \mathrm{DCR_R}=\frac{Vn_i }{2K_{\mathrm{gn}} }\cdot \phi,
    \label{eqndcrdose}
\end{equation}
where $V$ is the depletion region volume, $n_i$ is the depletion region intrinsic carrier density, $K_\mathrm{gn}$ is the material damage coefficient, and $\phi$ is the radiation dose. The damage coefficient $K_\mathrm{gn}$ measures radiation-induced defects in the depletion region and is independent of temperature, and we take $V$ and $K_{g_n}$ as constant~\cite{spadproperties}.

Displacement damage leads to the creation of intermediate energy levels in the semiconductor bandgap that allow thermal electrons to be more easily promoted to the conduction band, initiating an electron avalanche even in the absence of light. Such an anomalous event is considered a dark count. The dark count rate (DCR) generally increases with radiation dose, but the absolute rate cannot be predicted precisely because site damage is random. The general trend and variability in the dose response measured in radiation tests is shown in Supplementary Figure S9. The estimated DDD over the lifetime of SpooQy-1 is not expected to significantly change timing jitter, breakdown voltage, or detection efficiency that can affect quantum applications~\cite{cqtspaceborne,yang2019spaceborne}. Hence, we focus on $\dcrr$.

The SPEQS-2 instrument used the SAP500 model for GM-APDs, operated in free-running, passively quenched mode using a custom control circuit~\cite{chandrasekara15_spie}. This control circuit samples the avalanche pulse heights to dynamically control the applied voltage on the detector. This ensures that the avalanche pulse heights are kept significantly above the reference voltage of a discriminator that outputs digital pulses to an onboard counter. The other advantage of this circuit is that it maintains a constant excess voltage above the breakdown (hence constant detector efficiency) despite small drifts ($10^\circ$C) in temperature. This circuit also delays the onset of saturation behavior in the GM-APD, responding linearly to the increase in the photon rate up to 800,000 counts per second. Therefore, we expect that the reported DCR in this paper ($\approx$400,000 events per second) is still within the linear regime of the detector operation. For further GM-APD characteristics refer to Supplementary Note 2: GM-APD Characteristics.

\subsection*{Radiation modelling}

A representation of the radiation environment encountered must first be established. The expected space-flight radiation fluences are obtained from SPENVIS AP8/AE8 trapped particle models~\cite{spenvis_background} and chosen to be consistent with solar activity ~\cite{solarcycle25}, for solar and cosmic protons the SAPPHIRE and ISO15390 were chosen, respectively. SpooQy-1 CubeSat dropped from 418 km to 378 km (Fig.~\ref{DCR_altitudedata}(b).) between the deployment from the International Space Station until day 600 of the spaceflight. As radiation fluences are altitude and orbital trajectory dependent ~\cite{spaceraddirectionality,spenvis_background}, we accommodate this by simulating the radiation environment in 2.5 km increments in altitude drop along SpooQy-1's orbit. We import these radiation fluences from SPENVIS along with the CubeSat's CAD model (with material $Z$ atomic number and densities) into RSim. For each 2.5 km altitude drop we then simulate the associated dose rates. We combine this with the CubeSat telemetry to obtain the cumulative doses over time. The 2.5 km altitude increment provides an optimal compromise between the number of simulations required.  The benefit gradually diminishes for smaller increments. The contributions from each of the radiation sources (trapped electrons, trapped protons, solar protons, and cosmic protons) are simulated separately. The physics models and geometric biasing are optimized to provide a tolerable compromise between computation costs and low simulation errors (discussed in Supplementary Note 1: RSim modelling and parameters).

To correlate our radiation model with the accumulated radiation-induced dark count rates ($\dcrr$) we assume the following:

\begin{enumerate}\label{modelassumptions}
    \item No part-to-part variation in displacement damage-induced dark counts (that is, the constants in equation~\ref{eqndcrdose} are identical). 
    \item Negligible ionizing-dose-induced dark count rates of GM-APDs. The radiation tests reveal that displacement damage induced dark count rates exceed that produced by ionizing dose by several orders of magnitude ~\cite{sun01,yang2019spaceborne, cqtspaceborne} (refer to Supplementary Fig. S9).
    \item While the space-radiation environment has directional fluences ~\cite{spenvis_background, spaceraddirectionality,eastwesteffect}, because the satellite has no preferential orientation, i.e. it is designed to be tumbling while in orbit, it is safe to assume that it has received radiation fluences uniformly from all directions during its lifespan.
\end{enumerate}

The temperature differences between the thermistors near the two GM-APDs (see Fig.~\ref{APD_locations}) are minimal and within 3 degrees Celsius of each other (see Fig.~\ref{DCR_altitudedata}). Using the relationship between dark count rates and temperature ~\cite{sap500datasheet}, one can adjust them to the same temperature to compare dark counts. We note that spontaneous thermal annealing of radiation damage can take place, which may result in a lower $\dcrr$, but as the maximum temperature recorded within the satellite never exceeded $50^\circ$C, the temperature is too low to cause a significant recovery~\cite{lim2017laser} from the displacement damage~\cite{cqtspaceborne}. Temperatures below -20 degrees Celsius are also shown to reduce radiation damage ~\cite{apdthermalannealing1, apdthermalannealing2}, but these temperatures are not reached. Assumption 3 is motivated by the fact that the satellite is designed to be tumbling constantly in space, resulting in no preferential orientation for the spacecraft. Hence, the directionality associated with trapped particles arising from the East-West effect may be negligible ~\cite{spaceraddirectionality}.

Fig.~\ref{APD_locations} shows the 3D model that is imported into the RSim software for radiation analysis with the highlighted locations of the two GM-APDs. With RSim one can manually calculate the displacement damage dose in RSim by using the simulated spectral fluences into the GM-APDs and integrating with a Non-Ionzing-Energy-Loss (NIEL) calculator~\cite{spenvis_background, nielcalc}.

\subsection*{Data availability statement}
Data underlying the results presented in this paper are not publicly available at this time but may be obtained from the corresponding author upon reasonable request.

\subsection*{Code availability statement}
Code underlying the results presented in this paper are not publicly available at this time but may be obtained from the corresponding author upon reasonable request.

\subsection*{Competing interests}
The authors declare no competing interests

\subsection*{Author Contributions}
Arpad Lenart performed the radiation modelling for the mission and is the corresponding author for the manuscript. Tanvirul Islam worked on the design and construction of the SPEQS-2 and the SpooQy-1 and contributed to the writing of the manuscript. Srihari Sivasankaran conducted the flight data analysis and contributed to the writing of the manuscript. Peter Neilson provided assistance on radiation modelling in RSim. Bernhard Hidding and Daniel K.L. Oi supervised and contributed to the manuscript writing. Alexander Ling planned the SpooQy-1 mission and supervised the scientific studies.

\begin{acknowledgments}

DO acknowledges support from the EPSRC Researchers in Residence at the Satellite Applications Catapult (EPSRC Grant Ref: EP/T517288/1). DO and A. Lenart acknowledges support from the International Network in Space Quantum Technologies (EP/W027011/1), the EPSRC Quantum Technology Hub in Quantum Communication (EP/T001011/1), and the Integrated Quantum Networks Research Hub (EP/Z533208/1). A. Lenart acknowledges the support of an EPSRC Doctoral Training Partnership award. ALing acknowledges the support from the National Research Foundation, Singapore, under its Central Gap Fund (Ref: NRF2018NRFCG001-001).
\end{acknowledgments}

\newpage
\interlinepenalty=10000
\onecolumngrid 
\begin{center}
    \textbf{\large References}
\end{center}
\twocolumngrid 
\bibliographystyle{naturemag}
\providecommand{\noopsort}[1]{}\providecommand{\singleletter}[1]{#1}%

\clearpage
\newpage

\onecolumngrid 
\begin{center}
    \textbf{\large Supplementary Information: Comparing a radiation damage model for avalanche photodiodes through in-situ observation of CubeSat based devices}
\end{center}
\twocolumngrid 

\include{Supplementaryinfo_arxiv}

\end{document}

%% file: Supplementaryinfo_arxiv.tex
%
%
%
%
%
%
%

\renewcommand{\figurename}{Supp. Fig.}
\renewcommand{\thefigure}{S\arabic{figure}}

\renewcommand{\tablename}{Supp. Table.}

\title{Supplementary Information: Comparing a radiation damage model for avalanche photodiodes through in-situ observation of CubeSat based devices}

\maketitle


\subsection*{Supplementary Note 1: RSim modelling and parameters}\label{suppinforsimmodel}

RSim is a Monte-Carlo radiation modelling software based on the well established Geant4 radiation transport model. It allows 3D models to be imported and perform shielding calculations for geometries. By importing a particle energy spectrum into RSim one can evaluate the radiation effects by this field. To allow more accurate radiation transport modelling in well shielded areas geometric biasing is also available, whereby towards the center of the geometry of interest, the number of particles increase but their individual weights decrease.

After importing the 3D model of the CubeSat into RSim and the energy spectrum obtained from SPENVIS, we applied geometric biasing towards the center of the CubeSat to improve the modelling accuracy in the shielded regions where the GM-APDs are located. The physics models were set to "shielding physics" with options 4 and 3 for protons and electrons, providing a tolerable compromise between computational costs and simulation errors. Typically, Monte Carlo Radiation models require errors less than 10$\%$ error to be considered reliable.

\begin{figure}[H]
    \centering
    \includegraphics[width=\columnwidth]{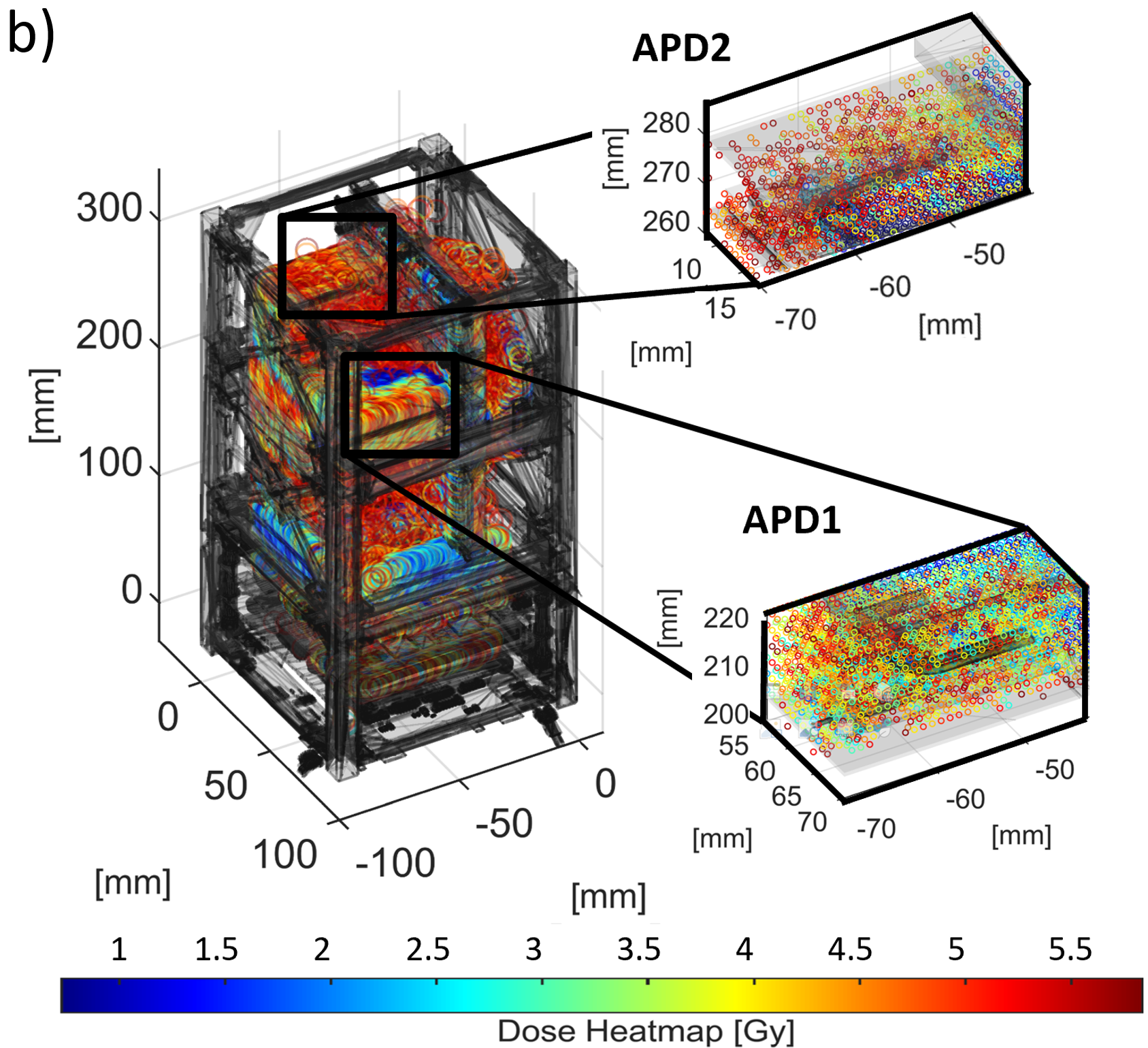}
    \caption{Heatmap of dose deposits by protons and electrons encountered by the SpooQy-1 CubeSat. The regions of the GM-APDs named ``APD1'' and ``APD2'' highlighted with two views. APD1 is near a dose coldspot,  and APD2 is more exposed to particles and is in a dose hotspot.}
    \label{APD_heatmapdose}
\end{figure}

Assuming a constant altitude of 410 km on average the yearly radiation doses into the two GM-APDs are shown in table ~\ref{apddoses} with a heatmap of total dose in ~\ref{APD_heatmapdose}.

\begin{table}
    \centering
   \begin{tabular}{|c|c|c|c|}
 \hline
 Particle & Energy [MeV] & APD1 [rads] & APD2 [rads] \\ [0.5ex] 
 \hline\hline\hline
 Electrons & & &\\
 \hline
 \hline
 Low Energy & 0-2 & 190.9 & 258.5\\
 \hline
 Medium Energy & 2-3 & 15.3 & 25.5\\
 \hline
 High Energy & 3-7 & 4.9 & 6.8 \\
 \hline\hline
 Protons & & \\
 \hline\hline
 Low Energy & 0-60 & 21.7 & 33.6 \\ [1ex] 
 \hline
 Medium  Energy & 60-200 & 18.0 & 23.8 \\ [1ex] 
 \hline
 High Energy & 200-400 & 3.5 & 3.8 \\ [1ex] 
 \hline
 Extreme Energy & 400-100,000 & 0.001 & 0.015 \\ [1ex] 
 \hline
 \hline
 Total Dose &  & 254.3 & 352.0\\ [1ex]
 \hline
    \end{tabular}
    \caption{Radiation dose contribution breakdown for an orbital trajectory consistent with the International Space Station at 410 km for a duration of 1 year). Detector doses from averaged omnidirectional particle fluences. For both GM-APDs, more than 99$\%$ of the dose is from trapped electrons and protons. Uncertainties in the total radiation doses are 5$\%$ 9$\%$ relative error for protons and electrons, respectively.}
    \label{apddoses}
\end{table}

\subsection*{Supplementary Note 2: GM-APD Characteristics}\label{suppinfoapdcharacteristics}

\begin{figure}[H]
    \centering
    \includegraphics[width=\columnwidth]{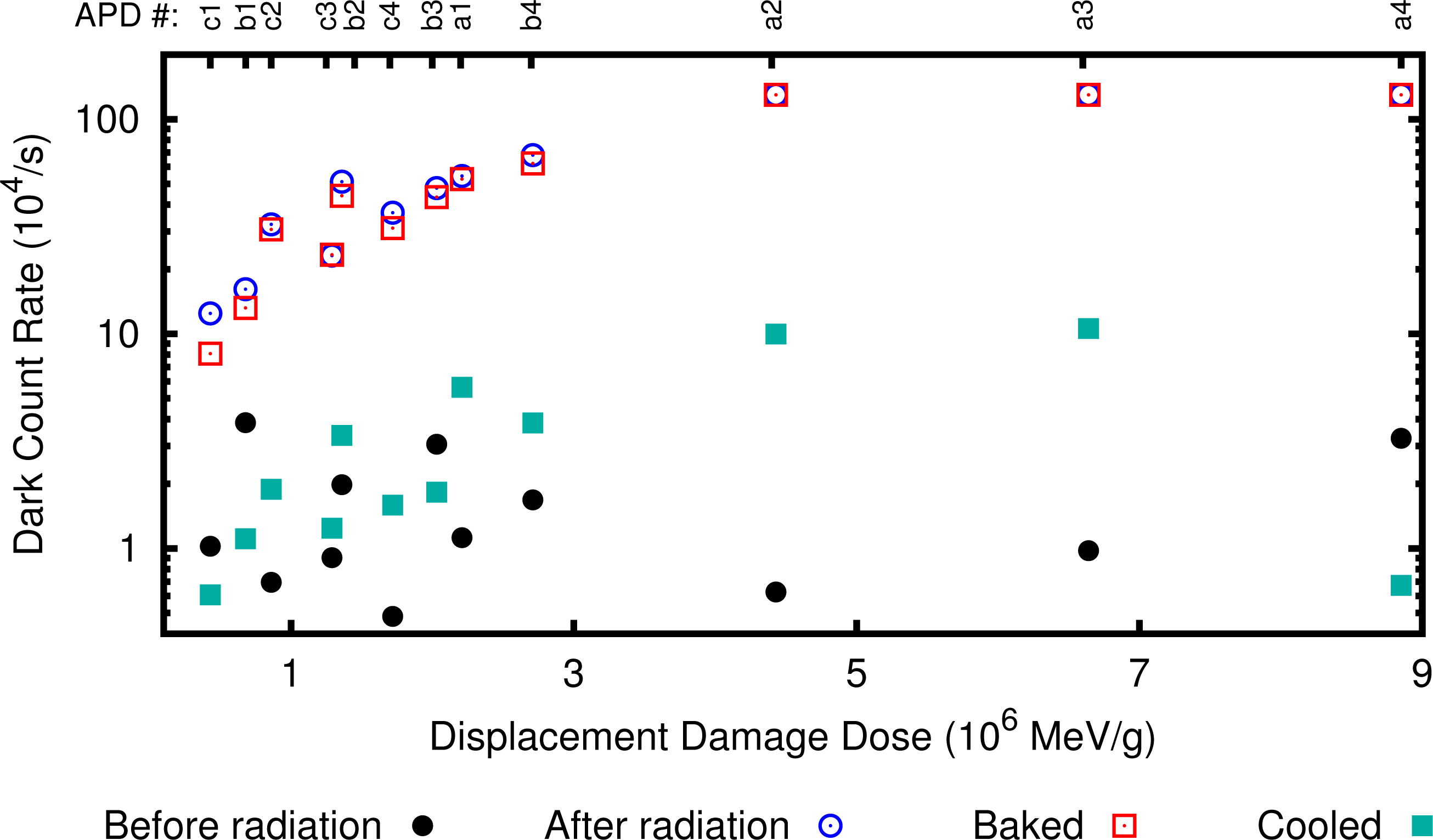}
    \caption{DCR vs DDD. Reproduced from \cite{cqtspaceborne}. GM-APD devices were exposed to different doses of displacement damage. A general correlation between displacement damage dose exposure and dark count rates were observed. Note that there is part-to-part variability in dark count response to radiation dose, but the general trend holds. The radiation tests were conducted by Crocker Nuclear Laboratory~\cite{cqtspaceborne} for 12 GM-APD devices with different proton energies (group a : 5 MeV, b : 25 MeV, c : 50 MeV) and 4 different total proton fluences were used for each group. The GM-APDs were operated at 22-25 degrees Celsius except when cooled to -20 degrees Celsius. It is noted that the four detectors with the largest DDD saturated the passive quench counting circuit, resulting in a plateau for detected dark counts at room temperature. Only a small recovery is observed after annealing (baking). Operating the detectors at low temperature (sub-zero Celsius) resulted in a massive reduction in dark counts, although typically still higher than pre-radiation rates.}
    \label{DCRddddependence}
\end{figure}

These are the operational parameters and characteristics for the SAP-500 GM-APDs used onboard SPEQS-2. For further details on the GM-APD SAP-500 consult the manufacturer's data sheet ~\cite{sap500datasheet}. The radiation test data from ~\cite{cqtspaceborne} is also shown in Fig. ~\ref{DCRddddependence}.

\centering
\begin{itemize}
    \item Average APD1 dark counts: 8,500 cps @10$~^\circ\text{C}$
    \item Average APD2 dark counts: 14,000 cps @10$~^\circ\text{C}$
    \item Detection efficiency: 0.55 at 780 nm and 0.40 at 850nm
    \item Working wavelength: 780 and 850 nm for APD1 and APD2, respectively.
    \item APD’s Operating mode: Free-running
    \item Dead-time: 1 $\mu$s
\end{itemize}


\subsection*{Supplementary Note 3: Dark count fitting parameters}

\begin{table}[H]
    \centering
    \begin{tabular}{|c|c|c c|c c|}
        \hline
        Day & & \multicolumn{2}{c|}{APD1} & \multicolumn{2}{c|}{APD2} \\ 
        \hline
        Count & & A & B & A & B \\ 
        \hline
        27 &  & 8.603587 & 0.084044  & 14.10942 & 0.084349  \\ 
        \hline
        190 &  & 18.46176 & 0.075322 & 51.57433 & 0.064257  \\ 
        \hline
        362 & & 47.71402 & 0.062356 & 70.03933 & 0.067188  \\ 
        \hline
        719 & & 45.75063 & 0.073212 & 90.52794 & 0.07293  \\ 
        \hline
    \end{tabular}
    \caption{Supplementary fitting parameters for the dark count rate trends in the GM-APDs for days 27, 190, 362, and 719. The fitting parameters are given in the form of $ DCR = A \cdot e^{B x} $}
    \label{apddcr_fitparameters}
\end{table}

